\documentstyle[12pt]{article}

\newfont{\set}{msbm10 scaled \magstep 1}
\newfont{\Set}{msbm6}

\newcommand{\be}{\begin{equation}}
\newcommand{\ee}{\end{equation}}
\newcommand{\bea}{\begin{eqnarray}}
\newcommand{\eea}{\end{eqnarray}}

\newcommand{\lag}{{\cal L}}

\newcommand{\var}{{\cal E}}

\def\pd{\partial}

\makeatletter

\def\eqnarray{\stepcounter{equation}\let\@currentlabel=\theequation
\global\@eqnswtrue
\global\@eqcnt\z@\tabskip\@centering\let\\=\@eqncr
$$\halign to \displaywidth\bgroup\@eqnsel\hskip\@centering
  $\displaystyle\tabskip\z@{##}$&\global\@eqcnt\@ne
  \hfil$\displaystyle{\hbox{}##\hbox{}}$\hfil
  &\global\@eqcnt\tw@ $\displaystyle\tabskip\z@
  {##}$\hfil\tabskip\@centering&\llap{##}\tabskip\z@\cr}
\@addtoreset{equation}{section}
  \def\theequation{\thesection.\arabic{equation}}
\makeatother

\begin{document}

\begin{titlepage}

\date{21 September 1995}
\title{BiHamiltonian Formulations of the Bateman Equation}

\author{J.A. Mulvey \\
        {\it Department of Mathematical Sciences} \\
        {\it University of Durham,} \\
        {\it South Road,} \\
        {\it DURHAM, DH1 3LE, UK} \\
        email: {\tt J.A.Mulvey@durham.ac.uk}}

\maketitle

{\centerline {\bf Preprint DTP/95/51}}

\begin{abstract}

  We discuss a class of evolution equations equivalent to the simplest
  Universal Field Equation, the so--called Bateman equation, and show
  that all of them possess (at least) biHamiltonian structure. The first
  few conserved charges are calculated.

\end{abstract}

\end{titlepage}

\section{Introduction}
Bateman's equation arose in a 1929 paper on hydrodynamics \cite{bate}
when he studied the properties of the equation:
\be
\Delta =
\phi_{x}^{2}\phi_{tt}-2\phi_{x}\phi_{t}\phi_{xt}+\phi_{t}^{2}\phi_{xx}=0.
\label{bateman}
\ee
Here $\phi$ is a scalar field and subscripts denote partial
derivatives. Many of the interesting properties of the Bateman
equation are dealt with in \cite{univ1}; its generalisations to higher
dimensions are discussed in that and subsequent papers
\cite{univ2,univ3,univ4}, where they become the {\it Universal Field
Equations}, a class of partial differential equations with intriguing
Lagrangian formulations.  Further analyis appears in \cite{grig}. In
the present case, we find that (\ref{bateman}) has an infinite number
of inequivalent Lagrangians. Any smooth, real--valued function
homogeneous of degree one in the derivatives $\phi_x$ and $\phi_t$,
and with arbitrary dependence on $\phi$, works as a Lagrangian for
$\Delta$.

This is straightforward to prove. Suppose $\lag(\phi,\phi_x,\phi_t)$
is a candidate Lagrangian. By Euler's theorem on homogeneous
functions:
\be
\phi_x \frac{\pd \lag}{\pd \phi_x}+\phi_t \frac{\pd \lag}{\pd
\phi_t}=\lag.
\label{eultheor}
\ee
It follows easily that:
\bea
\phi_x \frac{\pd^2 \lag}{\pd \phi_x^2}+\phi_t \frac{\pd^2 \lag}{\pd
\phi_x \pd \phi_t}&=&0, \nonumber \\
\phi_x \frac{\pd^2 \lag}{\pd \phi_x \pd \phi_t}+\phi_t \frac{\pd^2 \lag}{\pd
\phi_t^2}&=&0 \nonumber \\
\phi_x \frac{\pd^2 \lag}{\pd \phi \pd \phi_x}+\phi_t \frac{\pd^2
\lag}{\pd \phi \pd \phi_t}&=&\frac{\pd \lag}{\pd \phi}.
\label{hom0}
\eea
Now, if we expand the Euler--Lagrange expression,
\be
\var \lag = \frac{\pd \lag}{\pd \phi}-D_x \left (\frac{\pd \lag}{\pd
\phi_x} \right ) - D_t \left ( \frac{\pd \lag}{\pd \phi_t} \right ),
\label{eleq}
\ee
and use the homogeneity properties (\ref{hom0}) the equation of motion
becomes:
\be
\var \lag = \frac{1}{\phi_x \phi_t} \frac{\pd^2 \lag}{\pd \phi_x
 \pd \phi_t} \left ( \phi_{x}^{2}\phi_{tt} - 2\phi_{x}\phi_{t}\phi_{xt} +
\phi_{t}^{2}\phi_{xx} \right ).
\label{jp}
\ee
This is equivalent to (\ref{bateman}) on extremals of $\lag$.

Note that the homogeneous equations in (\ref{hom0}) imply that $\lag$
is always singular. Also, notice that if there is no $\phi$ dependence
in the Lagrangian, the Euler--Lagrange equation is in the form of a
conserved current, and so there are an infinite number of conservation
laws for (\ref{bateman}).  We will use this observation to construct
Hamiltonian conservation laws that are properly in involution.

It should be noted that much is already known about the Bateman
equation from the classical theory of partial differential equations.
For example, in common with other quasilinear second--order equations
\cite{c-h}, $\Delta$ can be linearised by a Legendre
transform \cite{univ4}. With the change of variables,
\bea
\xi &=& \phi_x \nonumber \\
\eta &=& \phi_t \nonumber \\
\omega \left ( \xi, \eta \right ) &=& x \xi +y \eta -  \phi,
\label{leg-trans}
\eea
(\ref{bateman}) is equivalent to:
\be
\xi^2 \omega_{\xi
\xi}+2 \xi\eta \omega_{\xi \eta}+\eta^2 \omega_{\eta \eta}.
\label{linbate}
\ee
This linearised equation supplies an implicit solution for
$\Delta=0$. According to a theorem of Kumei and Bluman
\cite{kum-blu}, the linearisability property is due to the existence
of a first--order generalised symmetry of (\ref{bateman}) whose
characteristic $Q$ depends on an arbitrary solution of the linear
equation:
\be
\phi_x^2 \frac{\pd^2 Q}{\pd \phi_x^2}+2\phi_x \phi_t
\frac{\pd^2 Q}{\pd \phi_x \pd \phi_t}+\phi_t^2 \frac{\pd^2
Q}{\pd \phi_t^2}=0.
\label{botheq-gs}
\ee
{}From this we deduce that any $Q$ homogeneous of degree zero or one in
the first derivatives of $\phi$ is a characteristic for a generalised
symmetry of (\ref{bateman}).

The task at hand is to study a more formal matter, namely whether
there is a Hamiltonian structure or structures associated with the
various Lagrangian formulations of the Bateman equation, and how its
integrability features in such a formalism. The basic reference for
the techniques used is \cite{olver}.

\section{Hamiltonian Analysis}

The Hamiltonian and multi-Hamiltonian descriptions of the Born--Infeld
equation have already been studied by Nutku and various collaborators
\cite{nut2,nut3} and the aim here is to perform a similar analysis of
the Bateman equation. The results of these papers can be adapted to
the present case simply by using the $\lambda=0$ limit of the
(unique) Born--Infeld Lagrangian,
\be
\lag_{BI}=\sqrt{\lambda +\phi_{x}^{2}-\phi_{t}^{2}}.
\label{b-i-lag}
\ee
This essentially repeats the work in \cite{nut3} but, as we
have seen, the set of admissible Lagrangians for (\ref{bateman}) is far
larger than this. We can discuss Hamiltonian formulations
corresponding to a large class of these non-standard Bateman
Lagrangians and in general we will find behaviour quite different from
the Born--Infeld case.

We need to find an evolution equation (or system of evolution
equations) of the form,
\be
{\bf u}_t={\cal J}\delta {\cal H}[{\bf u}],
\label{hamform}
\ee
where ${\cal H}[{\bf u}]$ is the Hamiltonian functional of the
dependent variables $u,v$ and ${\cal J}$
is a skew-adjoint differential operator. In this context,
``skew-adjoint'' means that
\be
\int_{\Omega} A.{\cal J}B dx=\int_{\Omega} B.{\cal J}^{*}A
dx=-\int_{\Omega} B.{\cal J}A dx,
\label{sk-a}
\ee
assuming that the support of $A$ and $B$ on the region $\Omega$ allows
formal integration by parts. The associated Poisson bracket is:
\be
\left \{ {\cal P},{\cal Q} \right \}
=\int \delta {\cal P}.{\cal J}\delta {\cal Q} dx.
\label{poisson}
\ee
This bracket must satisfy the Jacobi identity, which can be verified
using Olver's convenient method using functional tri-vectors
\cite{olver-jac,olver}. We require that this system is equivalent to
(\ref{bateman}) under some suitable change of variables.

Ideally, we would like to be able to find two such structures for our
evolution equations. Then, provided some compatibility criteria are
satisfied, Magri's theorem \cite{magri} guarantees the complete
integrability of the equation and allows construction of the Lax pair
representation and the associated conservation laws.

The most direct approach is suggested by the fact mentioned earlier,
that if the Bateman Lagrangian has no $\phi$ dependence,
(\ref{bateman}) can be expressed as a conservation law:
\begin{equation}
D_x \left ( \frac{\pd \lag}{\pd \phi_x} \right ) +D_t \left ( \frac{\pd
\lag}{\pd
\phi_t} \right )= 0.
\label{law}
\end{equation}
The biHamiltonian stucture of
equations expressible in the form of a conservation law was
demonstrated by Nutku \cite{nut1}. As mentioned above, there are an infinite
number of ways of expressing the Bateman equation as a divergence-free
current, and there is a biHamiltonian structure associated with each.

Following \cite{nut1}, the divergence (\ref{law}) may be rewritten as
the smoothness condition for a function $\psi$.
\bea
\psi_x & = & \frac{\pd \lag}{\pd \phi_t} \nonumber \\
\psi_t & = & \frac{\pd \lag}{\pd \phi_x}.
\label{psi}
\eea
In accordance with the standard Hamiltonian technique, the field
$\phi$ is mapped to a pair of variables $u,v$, where
\bea
u & = & \phi_x, \nonumber \\
v & = & \psi_x.
\label{new-fields}
\eea
The $t$--derivatives can then be expressed in terms of these new
variables:
\bea
\phi_t & = & U(u,v), \nonumber \\
\psi_t & = & V(u,v).
\label{t-deriv}
\eea
Consequently, the equation of motion and the smoothness condition for $\phi$
can be summarised by the condition of closure for the pair of exact $1$-forms,
\bea
d \phi & = & u\  dx + U\  dt, \nonumber \\
d \psi & = & v\  dx + V\  dt,
\label{exact}
\eea
namely that:
\bea
u_t-D_x \left ( U \right ) & = & 0 \nonumber \\
v_t-D_x \left ( V \right ) & = & 0.
\label{nut-evol}
\eea

Given our assumptions about the nature of the Lagrangian, we can
deduce the form of $U$ and $V$. From (\ref{hom0}), we know that the
derivatives of $\lag$ with respect to $\phi_x$ and $\phi_t$ are
homogeneous of degree zero. So, let us write (\ref{psi}) as,
\bea
\psi_x & = & F \left ( \frac{\phi_t}{\phi_x} \right ), \nonumber \\
\psi_t & = & G \left ( \frac{\phi_t}{\phi_x} \right ),
\label{soft-f}
\eea
where $F$ and $G$ are smooth, invertible functions. Solving the first
of these relations for $\phi_t$ tells us that,
\be
U(u,v)=u \ F^{-1}(v),
\label{Uexp}
\ee
and substitution into the second,
\be
V(u,v)=G \left ( F^{-1}(v) \right ).
\label{Vexp}
\ee

At this point, we attempt to impose the Hamiltonian structure. A nice
result about this type of equation \cite{nut1} is that it admits the canonical,
``flat'' Poisson structure defined by the structure matrix,
\be
{\cal J}_{1}=\left ( \begin{array}{cc}
                      0 & D_{x} \\
                      D_{x} & 0
                     \end{array} \right ),
\label{j1}
\ee
and a Hamiltonian ${\cal H}_1$ dependent only on $u$ and $v$, subject to the
conditions,
\be
\frac{\pd {\cal H}_{1}}{\pd u}=-V, \; \frac{\pd {\cal H}_{1}}{\pd v}=-U,
\label{h-compat}
\ee
which in turn impose compatibility criteria on $U$ and $V$.

In this example, the compatibility requires that:
\be
G^{\prime} \left ( F^{-1} (v) \right ) \frac{\pd}{\pd v} \left (
F^{-1} (v) \right ) = F^{-1}(v).
\ee
Bearing this in mind, we end up with a general Hamiltonian framework
for the Bateman equation. The Hamilton equations are easily seen to
reduce to the form,
\bea
u_t & = & D_x \left ( u F^{-1} (v) \right ), \nonumber \\
v_t & = & F^{-1} (v) \  v_x.
\label{ham-eq}
\eea
The structure matrix is ${\cal J}_1$ and the Hamiltonian is:
\be
{\cal H}_1=u \ G \left ( F^{-1} (v) \right ).
\label{1sthamgen}
\ee

There is also a biHamiltonian structure associated with these equations.
For this ``conserved current'' type of system there is always a
conserved quantity of the form \cite{nut1},
\be
{\cal H}_0=uv.
\label{h0}
\ee
Following the standard program for biHamiltonian systems \cite{olver},
we postulate this as a conserved Hamiltonian which reproduces the
equations of motion when used with a second structure matrix ${\cal
J}_2$:
\be
{\cal J}_1 \delta {\cal H}_1={\cal J}_2 \delta {\cal H}_0.
\label{2ham}
\ee
Provided this second structure is ``compatible'' with the first,
meaning that any linear combination of the two is also a Poisson
structure, this construction will give rise to a (hopefully infinite)
sequence of conserved quantities ${\cal H}_k$ such that,
\be
{\cal J}_1 \delta {\cal H}_k={\cal J}_2 \delta {\cal H}_{k-1},
\label{seqr}
\ee
which can be generated by the recursion operator:
\be
{\cal R}= {\cal J}_2 {\cal J}_1^{-1}.
\label{<rec-op}
\ee

We know that the general form of ${\cal J}_2$, as the most general
first--order, skew-adjoint, matrix differential operator is,
\be
{\cal J}_2=\left ( \begin{array}{cc}
                      2mD_x \ +m_x & 2pD_{x} \ +(p+q)_x \\
                      2pD_{x} \ + (p-q)_x & 2nD_x \ +n_x
                     \end{array} \right ),
\label{j2}
\ee
where $m,n,p,q$ are functions of $u$ and $v$. By using the condition
(\ref{2ham}) and demanding that it reproduces (\ref{ham-eq}),
restrictions on the form of ${\cal J}_2$ can be found. Nutku
\cite{nut1} proves that these restrictions also ensure the closure of
the Jacobi identity.

It is difficult to write down a convenient general form for the most
general structure so derived, but a useful simplifying assumption is
to take $n=0$. Then we find a possible candidate for the second
structure given by,
\bea
m & = & \frac{k}{v^2}, \nonumber \\
n & = & 0, \nonumber \\
p & = & \frac{1}{2}F^{-1}(v), \nonumber \\
q & = & \frac{1}{2}F^{-1}(v)+k^{\prime},
\label{2defs}
\eea
where $k$ and $k^{\prime}$ are constants. It is straightforward to
ascertain that this structure is compatible with the ${\cal J}_1$: the
combination ${\cal J}_1+{\cal J}_2$ is skew adjoint and satisfies the
Jacobi identity. So ${\cal J}_2$ can be used to find a second
conserved Hamiltonian and thence a whole hierarchy of conserved
quantities.

The associated recursion operator is the pseudodifferential operator,
\be
{\cal R}={\cal J}_2 {\cal J}_{1}^{-1}=\left ( \begin{array}{cc}
                      F^{-1}(v)+\frac{\pd}{\pd v}\left (F^{-1}\right
                      )v_x \ D_{x}^{-1} &
                      2\frac{k}{v^2}-2\frac{k}{v^3}v_x \ D_{x}^{-1} \\
                      0 & F^{-1}(v)
                     \end{array} \right ),
\label{gen-rec-op}
\ee
using the values of $p,q$, and $m$ in (\ref{2defs}).

It is known \cite{olver} that a system of evolution equations of the form,
\be
{\bf u}_t ={\bf K}[{\bf u}],
\ee
which admits a recursion operator, admits a Lax--type representation,
\be
{\sf  L}_t=\left [ {\cal A},{\sf L} \right ],
\label{lax}
\ee
if the ${\sf L}$ is the recursion operator (\ref{gen-rec-op}) and ${\cal
A}$ is the Fr{\'e}chet derivative:
\be
{\cal A}={\sf D}_{{\bf K}}=\left . \left (
\frac{d}{d \varepsilon} {\bf K}[{\bf u}_{\varepsilon}] \right ) \right
|_{\varepsilon = 0}.
\label{frech-def}
\ee
Here, ${\bf u}_{\varepsilon}$ denotes a one--parameter ($\varepsilon$) family
of
perturbations of ${\bf u}$.

In our example, the Fr{\'e}chet derivative is:
\be
{\cal A}=\left ( \begin{array}{cc}
   D_x \ \left ( F^{-1}(v)\ . \ \right ) & D_x \ \left ( u\frac{\pd
   }{\pd v}F^{-1} (v)\ . \ \right ) \\
                      0 & D_x \ \left ( F^{-1}(v)\ . \ \right )
                     \end{array} \right ).
\label{bate-frech}
\ee
The argument of the ${\cal A}$ should be inserted as indicated by the
dots. Substitution of (\ref{bate-frech}) and (\ref{gen-rec-op}) into
(\ref{lax}) reproduces the Hamilton equations (\ref{ham-eq}).

Now, using the recurrence (\ref{seqr}), we can start to calculate
conserved quantities. The first conserved Hamiltonian is found by
considering:
\be
{\cal J}_2 \delta {\cal H}_1={\cal J}_1 \delta {\cal H}_2.
\label{1st-cons}
\ee
It turns out to be:
\be
{\cal H}_2= \int^{v} \left \{ u\left ( F^{-1}(w)\right
)^2+\frac{2k}{w^2} G(F^{-1}(w)) \right \} dw .
\label{charge1}
\ee
Continuing the process, we find the third charge:
\be
{\cal H}_3 = \int^{v} \left \{ F^{-1}(w) \left (
u(F^{-1}(w))^2+\frac{2k}{w^2} G(F^{-1}(w))+\frac{2k}{w^2}
\int^{w}(F^{-1}( w^{\prime}))^{2} dw^{\prime} \right ) \right \} dw.
\label{charge2}
\ee
The subsequent members of the hierarchy become rapidly more
complicated.

It is worth noting that the conserved--current type of equation found
by Nutku admits a third independent Hamiltonian structure, compatible
with the first two, with ${\cal H}_2$ as the relevant Hamiltonian
density. Also, while the sequence of conserved charges in principle
continues {\it ad infinitum}, there may be certain particular cases
when the sequence terminates or repeats itself after a finite number
of steps. See \cite{nut3} for an example of this behaviour arising in
the biHamiltonian hierarchies of the Born--Infeld equation. Essentially
the same results can be found for the Bateman case by choosing the
Lagrangian (\ref{b-i-lag}) with $\lambda = 0$.

\vspace{1cm}

\noindent {\large {\bf Acknowledgements}} \newline
\noindent It is a pleasure to thank David Fairlie for his comments and
advice. The author is financially supported by the Department of
Education for Northern Ireland.

\end{document}